\documentclass[preprint,final,5p,times,twocolumn]{elsarticle}
\usepackage{rotating,color,subfigure,amssymb}
\usepackage{alphalph}
\usepackage{amsmath}
\usepackage[T1]{fontenc}
\usepackage{amssymb}
\journal{Physics Letters B}
\begin{document}

\begin{frontmatter}



\title{Measurement of the $np \to np\pi^0\pi^0$ Reaction in Search for the
  Recently Observed $d^*(2380)$ Resonance}

%\documentclass[final]{elsarticle}
%\begin{document}
%\begin{frontmatter}
%\title{}
\author[IKPUU]{The WASA-at-COSY Collaboration\\[2ex] P.~Adlarson\fnref{fnmz}}
\author[ASWarsN]{W.~Augustyniak}
\author[IPJ]{W.~Bardan}
\author[PITue,Kepler]{M.~Bashkanov}
\author[MS]{F.S.~Bergmann}
\author[ASWarsH]{M.~Ber{\l}owski}
\author[IITB]{H.~Bhatt}
\author[Budker,Novosib]{A.~Bondar}
\author[IKPJ,JCHP]{M.~B\"uscher\fnref{fnpgi,fndus}}
\author[IKPUU]{H.~Cal\'{e}n}
% ...

\address[IKPUU]{Division of Nuclear Physics, Department of Physics and 
 Astronomy, Uppsala University, Box 516, 75120 Uppsala, Sweden}
\address[ASWarsN]{Department of Nuclear Physics, National Centre for Nuclear 
 Research, ul.\ Hoza~69, 00-681, Warsaw, Poland}
\address[IPJ]{Institute of Physics, Jagiellonian University, ul.\ Reymonta~4, 
 30-059 Krak\'{o}w, Poland}
\address[PITue]{Physikalisches Institut, Eberhard--Karls--Universit\"at 
 T\"ubingen, Auf der Morgenstelle~14, 72076 T\"ubingen, Germany}
\address[Kepler]{Kepler Center for Astro and Particle Physics, Eberhard Karls 
 University T\"ubingen, Auf der Morgenstelle~14, 72076 T\"ubingen, Germany}
\address[MS]{Institut f\"ur Kernphysik, Westf\"alische Wilhelms--Universit\"at 
 M\"unster, Wilhelm--Klemm--Str.~9, 48149 M\"unster, Germany}
\address[ASWarsH]{High Energy Physics Department, National Centre for Nuclear 
 Research, ul.\ Hoza~69, 00-681, Warsaw, Poland}
\address[IITB]{Department of Physics, Indian Institute of Technology Bombay, 
 Powai, Mumbai--400076, Maharashtra, India}
\address[Budker]{Budker Institute of Nuclear Physics of SB RAS, 11~akademika 
 Lavrentieva prospect, Novosibirsk, 630090, Russia}
\address[Novosib]{Novosibirsk State University, 2~Pirogova Str., Novosibirsk, 
 630090, Russia}
\address[IKPJ]{Institut f\"ur Kernphysik, Forschungszentrum J\"ulich, 52425 
 J\"ulich, Germany}
\address[JCHP]{J\"ulich Center for Hadron Physics, Forschungszentrum J\"ulich, 
 52425 J\"ulich, Germany}
% ...

\fntext[fnmz]{present address: Institut f\"ur Kernphysik, Johannes 
 Gutenberg--Universit\"at Mainz, Johann--Joachim--Becher Weg~45, 55128 Mainz, 
 Germany}
\fntext[fnpgi]{present address: Peter Gr\"unberg Institut, PGI--6 Elektronische 
 Eigenschaften, Forschungszentrum J\"ulich, 52425 J\"ulich, Germany}
\fntext[fndus]{present address: Institut f\"ur Laser-- und Plasmaphysik, 
 Heinrich--Heine Universit\"at D\"usseldorf, Universit\"atsstr.~1, 40225 
 D??sseldorf, Germany}
\fntext[fnbe]{present address: Albert Einstein Center for Fundamental Physics,
 Universit\"at Bern, Sidlerstrasse~5, 3012 Bern, Switzerland}
\fntext[fnac]{present address: III.~Physikalisches Institut~B, Physikzentrum, 
 RWTH Aachen, 52056 Aachen, Germany}
\fntext[fnde]{present address: Department of Physics and Astrophysics, 
 University of Delhi, Delhi--110007, India}

%\end{frontmatter}
%\end{document}

\cortext[coau]{Corresponding author }

\begin{abstract}

Exclusive measurements of the quasi-free $np \to np\pi^0\pi^0$ reaction have
been performed by means of $dp$ collisions at $T_d$ = 2.27 GeV using the WASA
detector setup at COSY. Total and differential cross sections have been
obtained covering the energy region $\sqrt s$ = (2.35 - 2.46) GeV, which
includes the region of the ABC effect and its associated $d^*(2380)$
resonance.  
Adding the $d^*$ resonance amplitude to that for the conventional 
processes leads to a reasonable description of the data. 
The observed resonance effect in the total cross section is in agreement with
the predictions of F\"aldt and Wilkin as well Albadajedo and Oset.
The ABC effect, {\it i.e.} the low-mass enhancement in the
$\pi^0\pi^0$-invariant mass spectrum, is found to be very modest - if present
at all, which might pose a problem to some of its interpretations.
\end{abstract}

\begin{keyword}
Two-Pion Production, ABC Effect and Resonance Structure, Dibaryon Resonance
%Cross section, Double pion, Roper resonance
% MSC codes here, in the form: \MSC code \sep code
% or \MSC[2008] code \sep code (2000 is the default)

\end{keyword}
%\pacs{13.75.Cs, 14.20.Gk, 14.20.Pt}
\end{frontmatter}

% \linenumbers


%%%% typeset front matter (including abstract)

%%%%\maketitle 

\section{Introduction}

Recent data on the basic double-pionic fusion reactions $pn \to d\pi^0\pi^0$
and $pn \to d\pi^+\pi^-$ demonstrate that the so-called ABC effect is tightly
correlated with a narrow resonance structure in the total cross section of
these reactions \cite{prl2011,MB,isofus}. The ABC effect denoting a huge
low-mass enhancement in the $\pi\pi$ invariant mass spectrum is observed to
occur, if the initial nucleons or light nuclei fuse to a bound final nuclear
system and if the produced pion pair is isoscalar. The effect has been named
after the initials of Abashian, Booth and Crowe, who first observed it in the
inclusive measurement of the $pd \to ^3$HeX reaction more than fifty years ago
\cite{abc}. 

The resonance structure with $I(J^P) = 0(3^+)$ \cite{prl2011} observed in the
$pn \to d\pi\pi$ total cross section at $\sqrt s \approx$ 2.38 GeV is situated
about  80 MeV below $\sqrt s = 2 m_{\Delta}$, the peak position of the
conventional $t$-channel $\Delta\Delta$ process, which is also observed in
this reaction. The resonance structure has a width of only 70 MeV, which is
about three times narrower than the conventional process. From the Dalitz plot
of the $pn \to d\pi^0\pi^0$ reaction it is 
concluded that this resonance nevertheless decays via the intermediate
$\Delta^+\Delta^0$ system (at least predominantly) into its final
$d\pi^0\pi^0$ state. In the  $pn \to pp\pi^0\pi^-$ 
reaction the resonance has been sensed, too \cite{pp0-}, though in this case
there is no ABC effect associated with the resonance. In consequence it has no
longer be called ABC resonance, but $d^*$  -- adopting the notation of the
predicted so-called
"inevitable dibaryon" \cite{goldman} with identical quantum numbers.

By subsequent quasifree polarized $\vec{n}p$ scattering measurements it has been
demonstrated that there is a resonance pole in the coupled $^3D_3-^3G_3$
partial waves corresponding to the $d^*$ resonance structure in mass, width and
quantum numbers \cite{np,npfull} -- supporting thus its $s$-channel character. 

If this scenario is correct, then also the $np \to np\pi^0\pi^0$ reaction
should be affected by this resonance,  since this channel may proceed via the
same intermediate $\Delta^0\Delta^+$ system as the $np \to d \pi^0\pi^0$
and $pn \to pp\pi^0\pi^-$ reactions do. From a simple isospin point of view we
expect the resonance effect in the $np\pi^0\pi^0$ system to be identical in
size to that in the $d\pi^0\pi^0$ system. And from more refined estimates in
Refs. \cite{col,oset}, which account also for the different phase space
situations, we expect the resonance effect in
the $np\pi^0\pi^0$ channel to be about 85$\%$ of that in the $d\pi^0\pi^0$
system. Since the peak resonance cross section in the latter is 270 $\mu$b
\cite{isofus} sitting upon some background due to conventional $t$-channel
Roper and $\Delta\Delta$ excitations, we estimate the peak resonance
contribution in the $np\pi^0\pi^0$ system to be in the order of 200 $\mu$b. 

%Similar considerations apply also for the $np \to np\pi^+\pi^-$ reaction, as
%we will discuss at the end of this paper.

\section{Experiment}

Since there exist no data at all for the $np \to np\pi^0\pi^0$ channel, we
have investigated this reaction experimentally with the WASA detector at
COSY (FZ J\"ulich) by using a deuteron beam with an energy of
$T_d$~=~2.27~GeV impinging on a hydrogen pellet target \cite{barg,wasa}. By
exploiting the quasi-free scattering process $d p \to np\pi^0\pi^0 +
p_{spectator}$, we cover the full energy range of the conjectured resonance.
In addition, the quasi-free process in inverse kinematics gives us the
opportunity to detect also the fast spectator proton in the forward detector
of WASA.

The hardware trigger utilized in this analysis required at least two 
charged hits in the forward detector as well as two neutral hits in the central
detector.  

The quasi-free reaction $dp \to np \pi^0\pi^0 + p_{spectator}$
has been selected in the offline analysis by requiring two proton tracks in
the forward detector as 
well as four photon hits in the central detector, which can be traced back to
the decay of two $\pi^0$ particles. That way the non-measured neutron
four-momentum could be reconstructed by a kinematic fit with three
over-constraints.  

A difficulty emerges from deuterons, which originate from the $np \to
d\pi^0\pi^0$ reaction and which partly also break up while passing the detector. 
%Especially 
%those, which break up during passage through the 0.8 mm thick Aluminum window
%at the exit of the scattering chamber, 
Since in the $\Delta E-E$ energy loss plots used for particle identification
proton and deuteron bands overlap somewhat, deuterons can not be separated
completely from $np$ pairs
stemming from the $np \to np\pi^0\pi^0$ reaction. To suppress such misidentified
events we require the angle between emitted neutron and proton to be larger
than 5 degrees and also their energies to be in the expected
range. Nevertheless a Monte-Carlo (MC) simulation of 
the  $np \to d\pi^0\pi^0$ reaction, which is known in very detail
\cite{prl2011}, shows that we have to expect still a contamination of about
5$\%$ in the spectra of the $np \to np\pi^0\pi^0$ reaction. In Figs. 1 - 6 the
observables are shown with the MC-generated contamination events already
subtracted. In the $pn$ 
invariant-mass spectrum $M_{pn}$, where the contamination shows up most
pronounced, this concerns only the first two bins (Fig.~3, bottom).     

In Fig. 1
the measured and acceptance corrected spectator momentum distribution is shown
in comparison with a 
Monte-Carlo (MC) simulation of the quasifree $dp \to np \pi^0\pi^0 +
p_{spectator}$ process. Due to the beam-pipe ejectiles can only be detected in
the WASA forward detector for lab angles larger than three degrees. The good
agreement between data and simulation provides confidence that the data indeed
reflect a quasifree process.  
%As in Refs. \cite{prl2011,isofus,np} we use only spectator momenta
%$p_{spectator} <$ 0.16 GeV/c for the further data analysis. 
The constraint for the suppression of breakup events (see above) causes the
maximum accepted spectator momentum to be $<$ 0.14 GeV/c fulfilling the
spectator momentum condition used in previous works  \cite{prl2011,isofus,np} 
This implies 
an energy range of 2.35 GeV $\leq \sqrt s \leq$ 2.41 GeV being covered due to
the Fermi motion of the nucleons in the deuteron. This energy range
corresponds to incident lab energies of 1.07 GeV $< T_n < $ 1.23 GeV.

\begin{figure} 
\centering
\includegraphics[width=0.89\columnwidth]{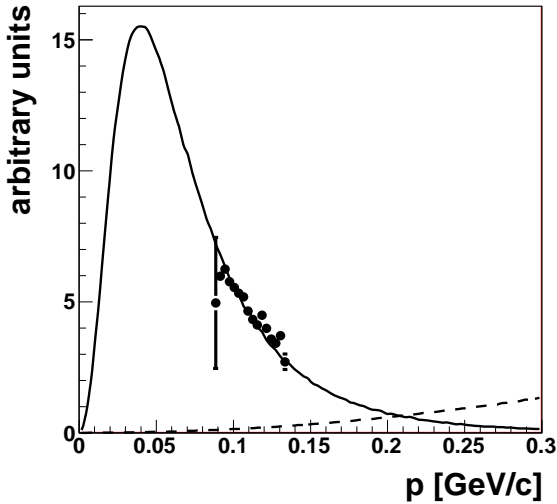}
\caption{\small
  Efficiency corrected distribution of the spectator proton momenta in the $dp
  \to np\pi^0\pi^0 +  p_{spectator}$ reaction within the WASA acceptance,
  which allows the detection of the spectator proton only for lab angles
  larger than three degrees. In addition the constraint for the suppression of
  breakup events has been applied (see text).
  Data are given by solid circles. The solid line shows 
  the expected distribution for the quasifree process based on the CD Bonn 
  potential \cite{mach} deuteron wavefunction. For comparison the dashed line
  gives the pure phase-space distribution as expected for a coherent reaction
  process.
}
\label{fig1}
\end{figure}

In total a sample of about 24000 good events has been selected.
The requirement that the two protons have to be in the angular range covered
by the forward detector and that the gammas resulting from
$\pi^0$ decay have to be in the angular range of the central detector reduces
the overall acceptance to about 7$\%$. 
Efficiency and acceptance corrections of the data have been performed by MC
simulations of reaction process and detector setup. For the MC simulations
model descriptions have been used, which will be discussed in the next
chapter. Since the acceptance is substantially below 100$\%$, the efficiency
corrections are not fully model independent. The hatched grey histograms in 
Figs. 3 - 6 give an estimate for systematic uncertainties due to the use of
different models with and without $d^*$ resonance hypothesis for the
efficiency correction.  

The absolute normalization
of the data has been performed by the simultaneous measurement of the
quasi-free single pion production process $dp \to pp \pi^0 + n_{spectator}$
and its comparison to previous bubble-chamber results for the $pp
\to pp \pi^0$ reaction \cite{shim,eis}. That way the uncertainty in the
absolute normalization of our data is essentially that of the previous $pp \to
pp \pi^0$ data, {\it i.e.} in the order of 20$\%$.

\section{Results and Discussion}

In order to determine the energy dependence of the total cross section we have
divided our data sample into 10 MeV bins in $\sqrt s$. The resulting total
cross sections together with their statistical and systematic uncertainties
are listed in Table 1.

\begin{table}
\caption{Total cross sections obtained in this work for the $np \to
  np\pi^0\pi^0$ reaction in dependence of the center-of-mass energy $\sqrt s$
  and the neutron beam energy $T_n$. Systematic uncertainties are given as
  obtained from MC simulations for the detector performance assuming various
  models for the reaction process. } 
\begin{tabular}{llllll} 
\hline

 & $\sqrt s$ & $T_n$ &~~~~$\sigma_{tot}$  &~~~~$\Delta\sigma_{stat}$ &~~~~$\Delta\sigma_{sys}$  \\ 
& [MeV] & [MeV] &~~~~[$\mu$b] &~~~~[$\mu$b]  &~~~~[$\mu$b] \\

\hline

& 2.35 & 1.075 &~~~~127 &~~~~6 &~~~12 \\
& 2.36 & 1.100 &~~~~192 &~~~~9 &~~~20  \\
& 2.37 & 1.125 &~~~~222 &~~~11 &~~~22  \\
& 2.38 & 1.150 &~~~~269 &~~~13 &~~~27  \\ 
& 2.39 & 1.176 &~~~~293 &~~~14 &~~~29  \\ 
& 2.40 & 1.201 &~~~~295 &~~~14 &~~~29  \\
& 2.41 & 1.227 &~~~~272 &~~~13 &~~~27 \\

\hline
 \end{tabular}\\
\end{table}

Fig.~2 exhibits the energy dependence of the total cross section for the $np
\to np\pi^0\pi^0$ reaction (right) in comparison to that of the $pp \to
pp\pi^0\pi^0$ reaction (left). The previous WASA results \cite{iso,TT} and the
ones of this work are given by the full circles. They are compared to previous
bubble-chamber measurements from KEK (open circles) \cite{shim} in case of the
$pp\pi^0\pi^0$ channel. 

In case of the $np\pi^0\pi^0$ channel there exist no
dedicated data from previous investigations. However, there are some connected
data from the PINOT experiment at Saclay, where the inclusive reactions $pp
\to \gamma\gamma X$ and $pd \to \gamma\gamma X$ were measured at $T_p$ = 1.3
and 1.5 GeV \cite{Scomparin}. By excluding the two-photon invariant mass regions
corresponding to single $\pi^0$ or $\eta$ production the remaining two-photon
events populating the combinatorial background are likely to originate 
from $\pi^0 \pi^0$ production. By using this feature a measure of the ratio
of the cross sections $pn \to pn\pi^0\pi^0 + d\pi^0\pi^0$ to $pp \to
pp\pi^0\pi^0$ has been obtained. This leads to a crude estimate for the  $pn \to
pn\pi^0\pi^0$ cross section to be larger than the $pp \to pp\pi^0\pi^0$ cross
section by roughly a factor of two -- in qualitative support of our results
from the exclusive measurements \cite{Colin}.

\begin{figure*} [t]
\begin{center}
\includegraphics[width=0.99\textwidth]{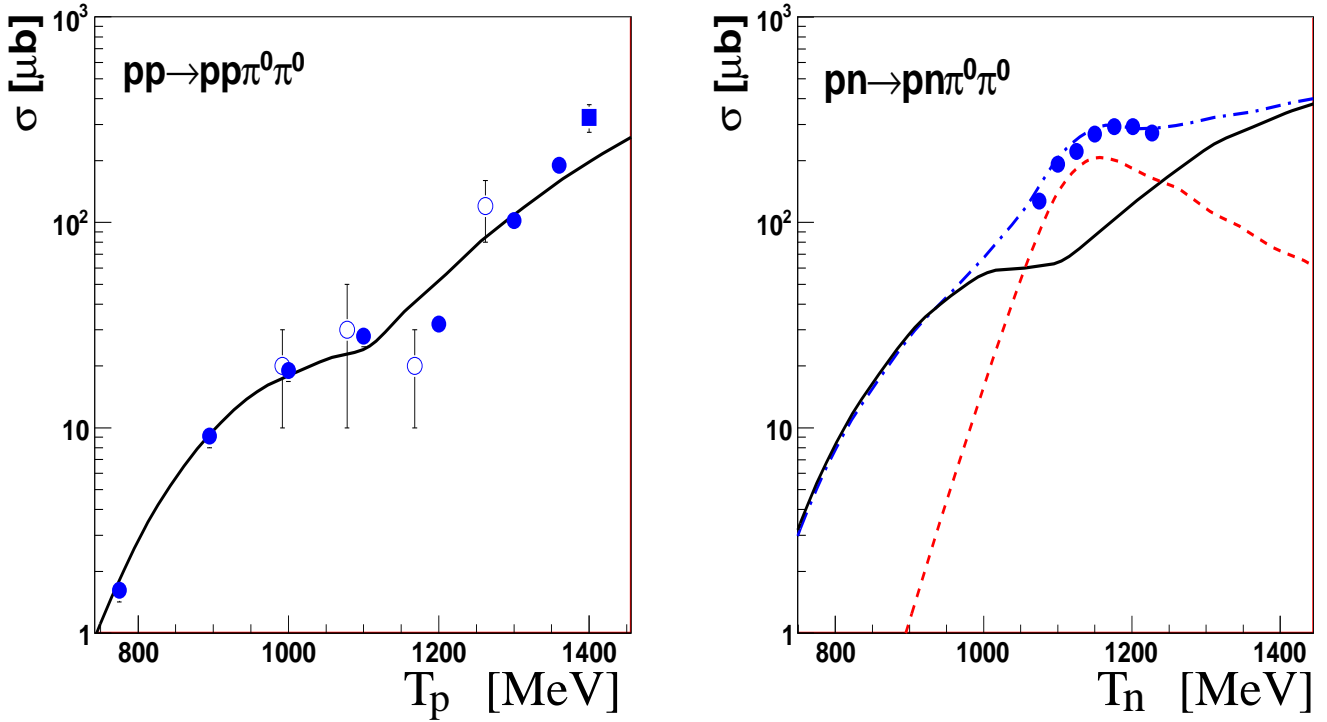}
\caption{(Color online) 
   Total cross sections for the reactions $pp \to pp\pi^0\pi^0$ (left) and $np
   \to np\pi^0\pi^0$ (right). The results of this work are shown by the full
   circles in the right figure. Previous WASA results on the
   $pp\pi^0\pi^0$ channel are shown by full circles \cite{iso} and full
   square \cite{TT}, respectively, in the left figure,
   previous bubble-chamber measurements from 
  KEK \cite{shim} by open circles. The modified
  Valencia model calculation is shown by the solid lines. The dash-dotted
  curve shows the result, if the $s$-channel $d^*$ resonance amplitude is
  added. The $d^*$ contribution itself is given by the dotted curve.
}
\label{fig2}
\end{center}
\end{figure*}

In Fig.~2 we compare the data to theoretical calculations in the framework of
the Valencia model \cite{luis}, which incorporates both non-resonant and
resonant $t$-channel processes for two-pion production in $NN$
collisions. The $t$-channel resonance processes of interest here concern first
of all the
excitation of the Roper resonance and its subsequent decay 
either directly into the $N\pi\pi$ system or via the $\Delta\pi$
system as well as the excitation and decay of the $\Delta\Delta$
system. Deviating from the original Valencia calculations \cite{luis} the
present 
calculations have been tuned to describe quantitatively the isovector two-pion
production reactions $pp \to NN\pi\pi$ \cite{iso}, in particular the
$pp\pi^0\pi^0$ \cite{deldel} and $nn\pi^+\pi^+$ \cite{nnpipi} channels by the
following modifications: 

\begin{itemize}
\item relativistic corrections for the $\Delta$ propagator as given by
  Ref.~\cite{ris},
\item strongly reduced $\rho$-exchange contribution in the $t$-channel
  $\Delta\Delta$ process -- in agreement with calculations from Ref.~\cite{xu},
\item reduction of the $N^* \to \Delta\pi$ amplitude by a factor of two in
  agreement with the analysis of photon- and pion-induced pion production on
  the nucleon \cite{boga} and in
  agreement with $pp \to pp\pi^0\pi^0$ and $pp \to pp\pi^+\pi^-$ measurements
  close to threshold \cite{WB,JP,ae,Roper} as well as  readjustment of the
  total Roper excitation according to the results of the isospin decomposition
  of the $pp \to NN\pi\pi$ cross sections \cite{iso}, 
\item inclusion of the $t$-channel excitation of the $\Delta(1600)P_{33}$
  resonance.
\end{itemize}

The latter modification was necessary, in order to account for the
unexpectedly large $pp \to nn\pi^+\pi^+$ cross section \cite{nnpipi}. The
predictive power of these modifications has been demonstrated by its
successful applications to the recent $pp \to pp\pi^0\pi^0$ data at $T_p$ =
1.4 GeV \cite{TT} and to the $pn \to pp\pi^0\pi^-$ reaction \cite{pp0-}. 

Final state interaction (FSI) in the emitted $NN$ system has been taken into
account in the Migdal-Watson \cite{migdal,watson} factorized form.
% though in the
%energy region of interest here -- {\it i.e.} 1.1 $\lesssim T_{lab}
%\lesssim$ 1.4 GeV, which is already substantially above the $\pi\pi$
%production threshold -- the $NN$ final state interaction is not of crucial
%importance. 
%In case of the $pp \to
%pp\pi^0\pi^0$ reaction at 1.3 GeV this has been demonstrated in Ref.
%\cite{deldel} by means of the 
%$M_{pp}$ spectrum (Fig.~6, top left in Ref. \cite{deldel}), where the solid
%line shown there exhibits only a small enhancement at threshold due to the $pp$
%final state interaction. 

The $NN$ FSI is by far strongest in the isovector $^1S_0$
$pn$ state and less strong in $^1S_0$ $pp$ and $^3S_1$ $pn$ states as apparent
from the scattering lengths in these systems. At energies above 1 GeV the
$t$-channel $\Delta\Delta$ process is the dominating one. Isospin
decomposition of its contribution to the total $np \to np\pi^0\pi^0$ cross
section \cite{dakhno,bys,iso} shows that in this process the $^1S_0$ final
state is much less populated than the isoscalar $^3S_1$  state. The situation
is somewhat different in the near-threshold region, where the Roper excitation
process dominates. In this process equal amounts of $pn$ pairs are emitted in
$^1S_0$ and $^3S_1$ states. 
 
Since the modified Valencia calculations have been tuned to the $pp \to
pp\pi^0\pi^0$  reaction, it is no surprise that its total cross section is
fairly well described -- see Fig. 2, left. For the closely related $np \to
np\pi^0\pi^0$ reaction the calculations predict a similar energy dependence,
but an absolute cross section, which is larger by roughly a factor of two --
whereas the data are larger by more than an order of magnitude -- see Fig. 2,
right. 

As an independent check of these calculations we may perform an isospin
decomposition of cross sections using the formulas given in
Refs. \cite{dakhno,bys} and the matrix elements deduced from the
analysis of the $pp$ induced two-pion production \cite{iso}. As an result of
such an exercise we get agreement with the modified Valencia calculation within
roughly 30$\%$. 
  
As we see from Fig.~2, the experimental cross sections obtained in this work
for the $np \to np\pi^0\pi^0$ reaction are three to four times larger than
predicted.  This failure points to an important reaction component not
included in the $t$-channel treatment of two-pion production. It is intriguing
that we deal here with the energy region, where the $d^*$ resonance has been
observed both in $np$ scattering \cite{np} and in the isoscalar part of the
double-pionic fusion to deuterium \cite{prl2011,isofus}. Also it has been
shown that the description of the $pn \to pp\pi^0\pi^-$ cross section improves
greatly in this energy region, if this resonance is included
\cite{pp0-}. Hence we add also here the amplitude 
of this resonance to the conventional amplitude. According to the predictions
of F\"aldt and Wilkin \cite{col} as well as Abaladejo and Oset \cite{oset},
its contribution at the resonance maximum should be about 200 $\mu$b (dotted
curve in 
Fig.~2) as discussed in the introduction. It is amazing, how well the
resulting curve (dash-dotted line in Fig.2) describes the data. Of course, it
is a pity that there are no data outside the energy region covered by our
data. In particular at energies below 1 GeV and above 1.3 GeV, {\it i.e.}
outside the resonance region, such data would be very helpful to examine
experimentally the reliability of the predictions for the $t$-channel
contributions.  
%Adjusting the
%resonance contribution to the data requires a peak cross section in the
%range of 90 - 130 $\mu$b -- depending on the systematic
%uncertainties associated with our values for the total cross section.

When binned into $\sqrt s$ bins of 10 MeV the different distributions do not
exhibit any particular energy dependence in their shapes -- which is of no
surprise, since the energy region covered in this measurement is dominated by
the $d^*$ resonance as evident from the discussion of the total cross
section. Hence we refrain from showing 
the differential distributions for single $\sqrt s$ bins. We rather
show them unbinned, {\it i.e.}, averaged over the full energy range of the
measurement, which has the advantage of better statistics and less systematic
uncertainties.

For a four-body final state there are seven independent differential
observables. We choose to show in this paper the differential distributions
for the invariant masses $M_{\pi^0\pi^0}$, $M_{pn}$, $M_{p\pi^0}$,
$M_{n\pi^0}$, $M_{n\pi^0\pi^0}$ and $M_{pp\pi^0}$ as well as the differential
distributions for the center-of-mass (cm) angles for protons and pions, namely
$\Theta_p^{c.m.}$ and $\Theta_{\pi^0}^{c.m.}$. These distributions are shown in
Figs. 3 - 6. 
%with each of them plotted for four ?????
%energy bins: 2.35 GeV $ < \sqrt s <$ 2.36 GeV (a), 2.365 $< \sqrt s <$ 2.375
%GeV (b), 2.40 $< \sqrt s <$ 2.41 GeV (c) and 2.44 GeV $< \sqrt s <$ 2.45 GeV
%(d) ???????. 
%These bins are denoted in the figures by their central values $\sqrt s$ =
%2.355, 2.370, 2.405 and 2.445 GeV, respectively. 
%The second region is chosen to cover just the peak region of the $d^*$
%resonance structure observed in the $pn \to d\pi^0\pi^0$ reaction. ?????

\begin{figure} %[t]
\begin{center}
%\parbox[c] {0.49\textwidth}{
%\centering

\includegraphics[width=0.4\textwidth]{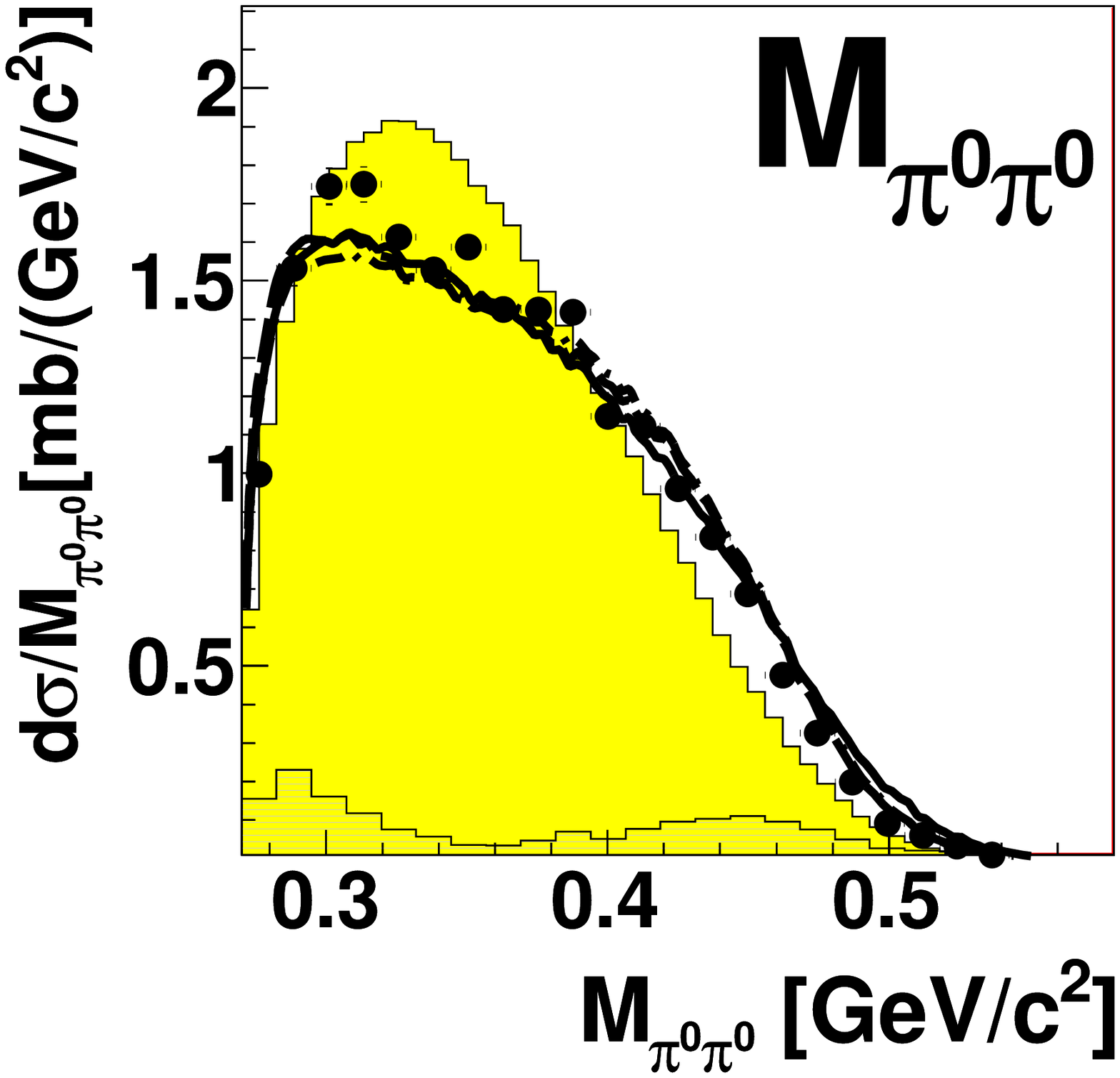}
\includegraphics[width=0.4\textwidth]{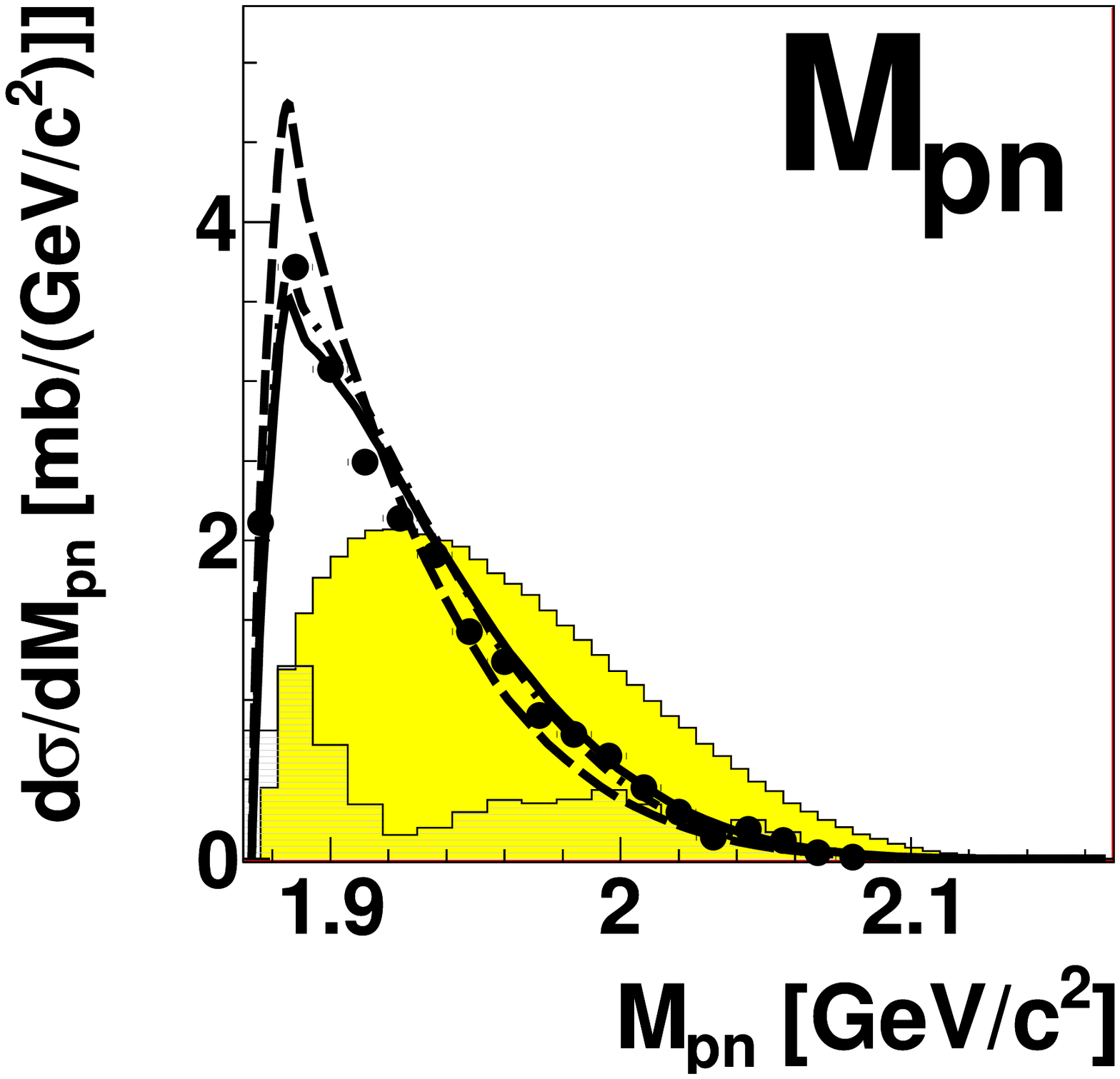}

\caption{(Color online) Top:
   distribution of the $\pi^0\pi^0$ invariant mass $M_{\pi^0\pi^0}$ for the $pn
   \to np\pi^0\pi^0$ reaction at $T_n$ = 1.135 GeV. Since the data are shown
   without separation into $\sqrt s$ bins, they correspond to the 
   average over the energy region covered by the quasifree collision process,
   which is 2.35 GeV $ < \sqrt s <$ 2.41 GeV (1.07 GeV $< T_n <$ 1.23 GeV).  
   Filled
   circles represent the experimental results of this work. The hatched
   histograms give estimated systematic uncertainties due to the incomplete
   coverage of the solid angle. The shaded
   areas denote phase space distributions. The solid lines are calculations 
   with the modified Valencia model. The dashed (dash-dotted) lines shows the
   result, if the $d^*$ resonance amplitude with (without) inclusion of the
   $\Delta\Delta$ vertex function \cite{prl2011} is added. 
%   The dash-dotted lines give the calculations, where the vertex function is
%   replaced by a $d^* \to d\sigma$ contribution according to the ansatz of
%   Ref.~\cite{kuk}. 
   All calculations are normalized in area to the data. 
Bottom: the same as at the top, but for the $pn$ invariant mass $M_{pn}$.
}
\label{fig3}
\end{center}
\end{figure}

\begin{figure} [t]
\begin{center}

\includegraphics[width=0.4\textwidth]{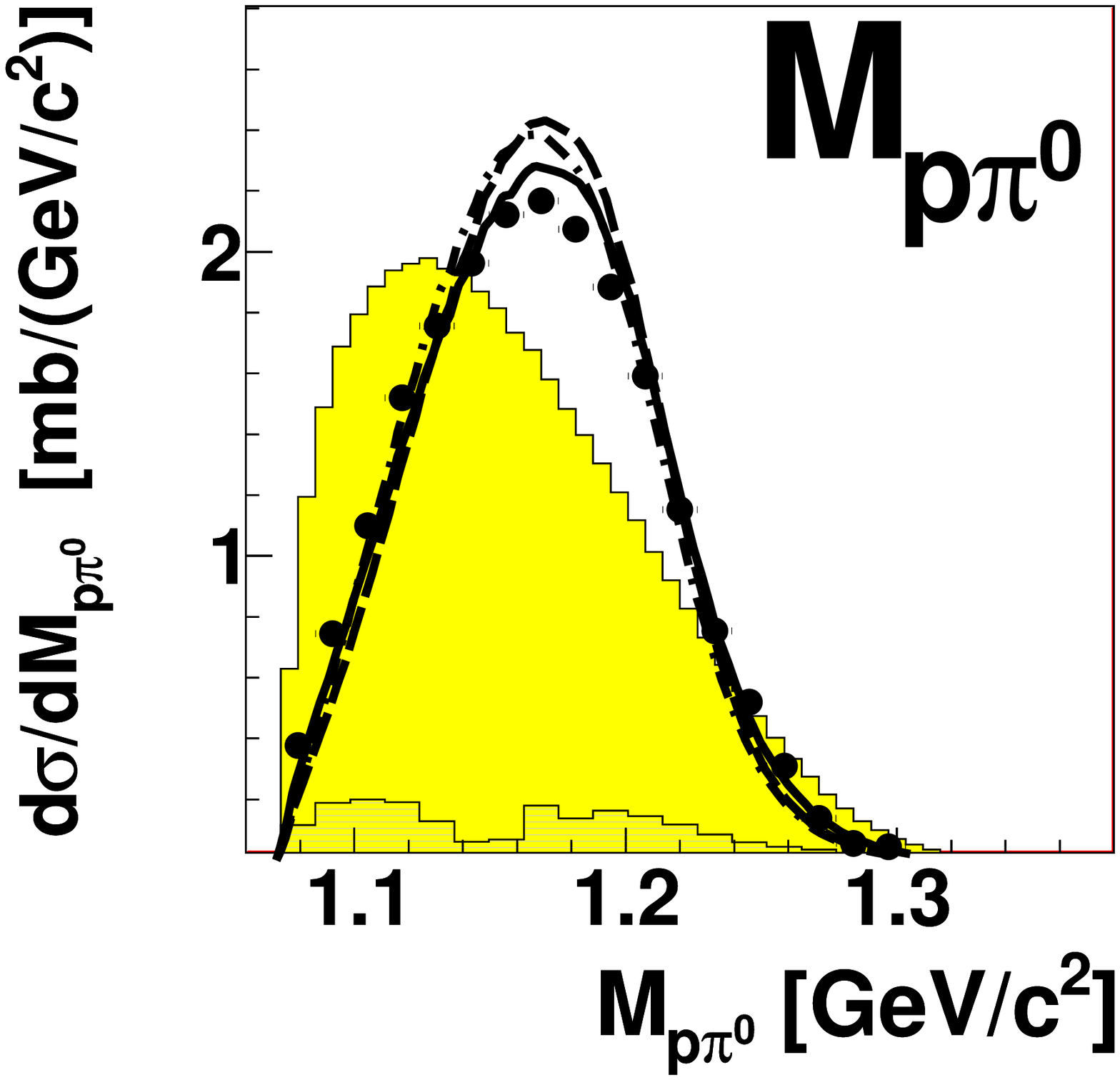}
\includegraphics[width=0.4\textwidth]{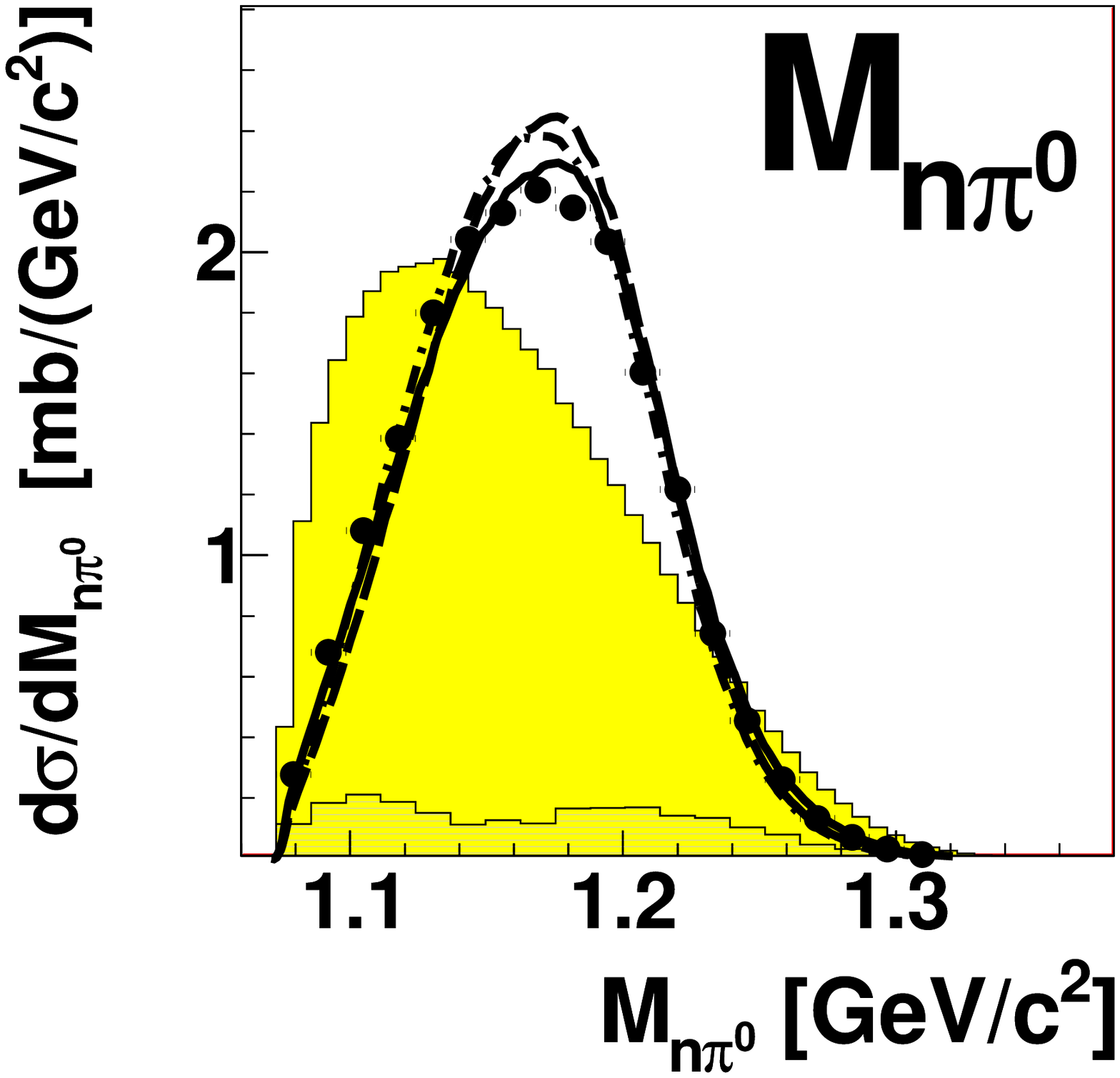}

\caption{(Color online) 
  Same as Fig. 3 but for the 
   distributions of the invariant masses $M_{p\pi^0}$ (top) and $M_{n\pi^0}$
   (bottom). 
}
\label{fig4}
\end{center}
\end{figure}

\begin{figure} [t]
\begin{center}
\includegraphics[width=0.4\textwidth]{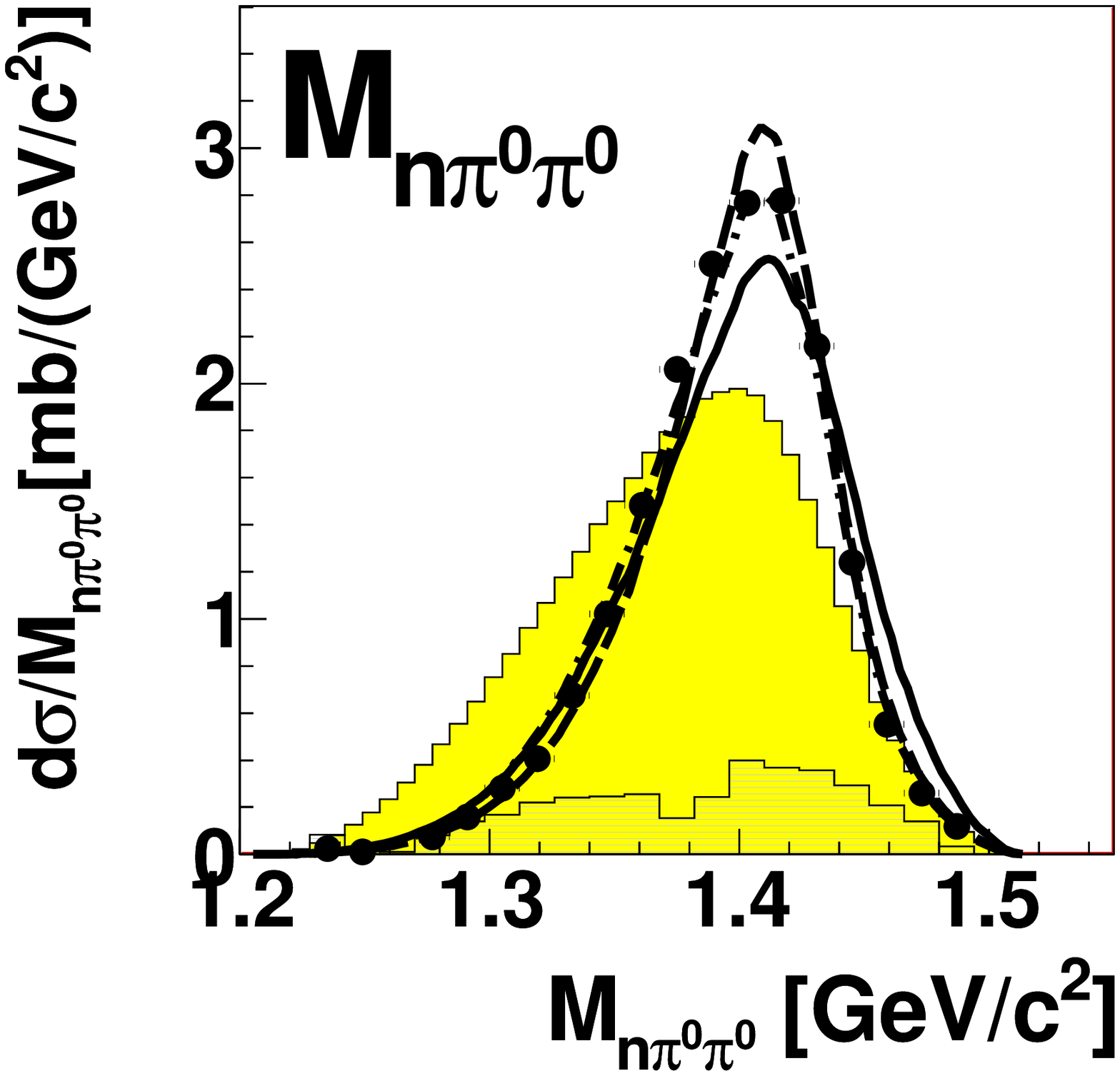}
\includegraphics[width=0.4\textwidth]{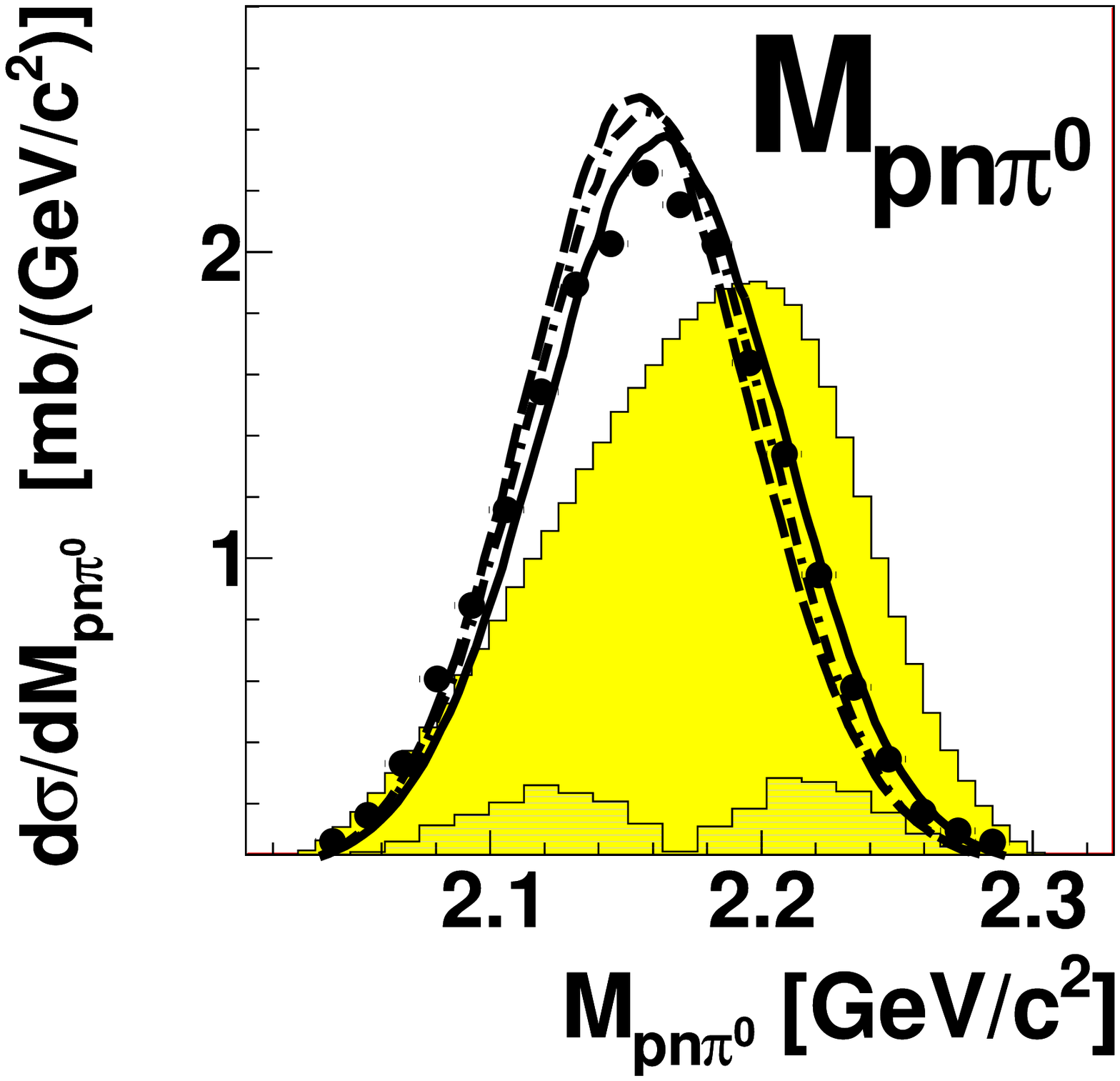}
\caption{(Color online) 
  Same as Fig. 3 but for the 
   distributions of the invariant masses $M_{n\pi^0\pi^0}$ (top) and
   $M_{pn\pi^0}$ (bottom).  
}
\label{fig5}
\end{center}
\end{figure}

\begin{figure} [t]
\begin{center}
\includegraphics[width=0.4\textwidth]{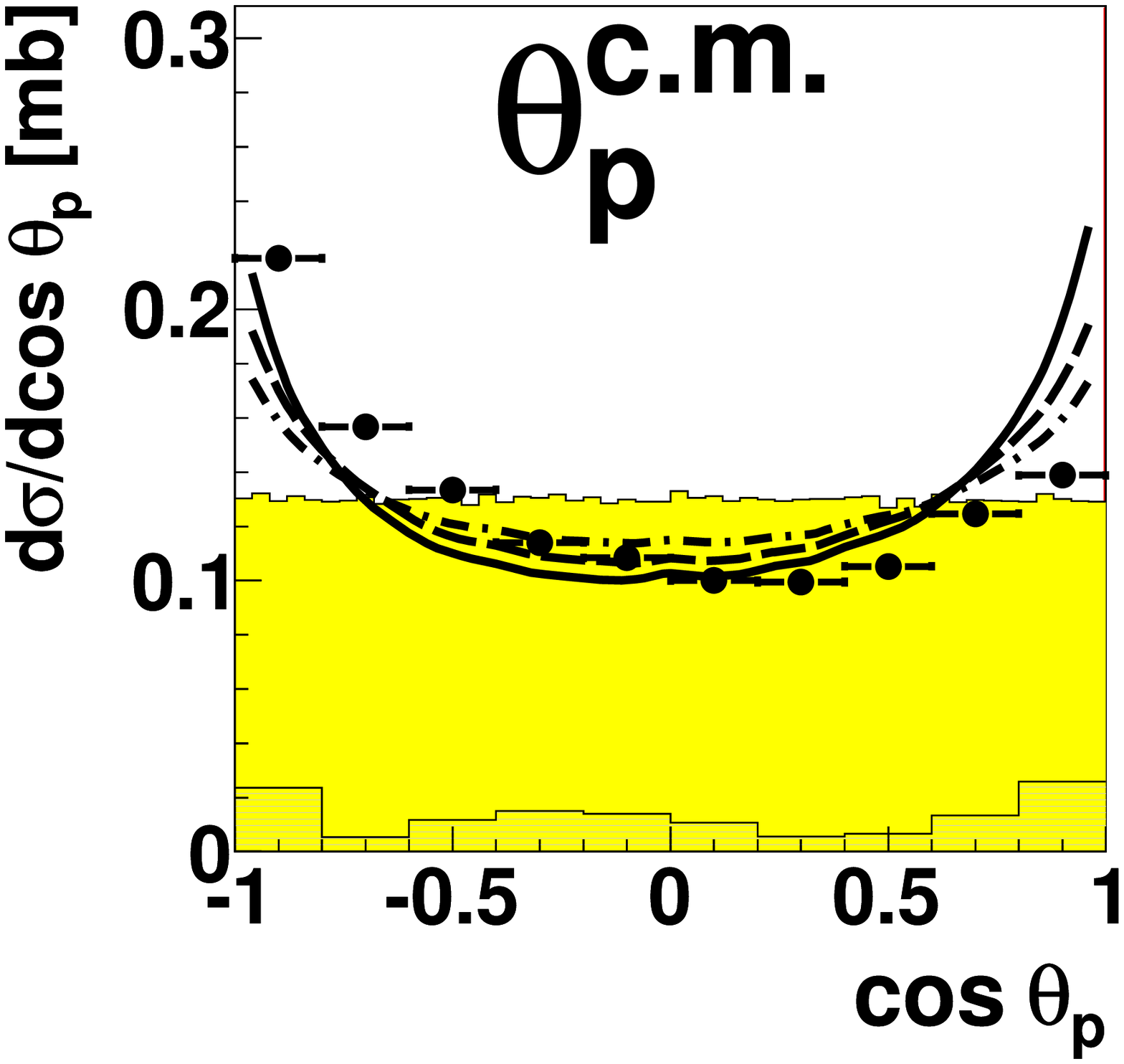}
\includegraphics[width=0.4\textwidth]{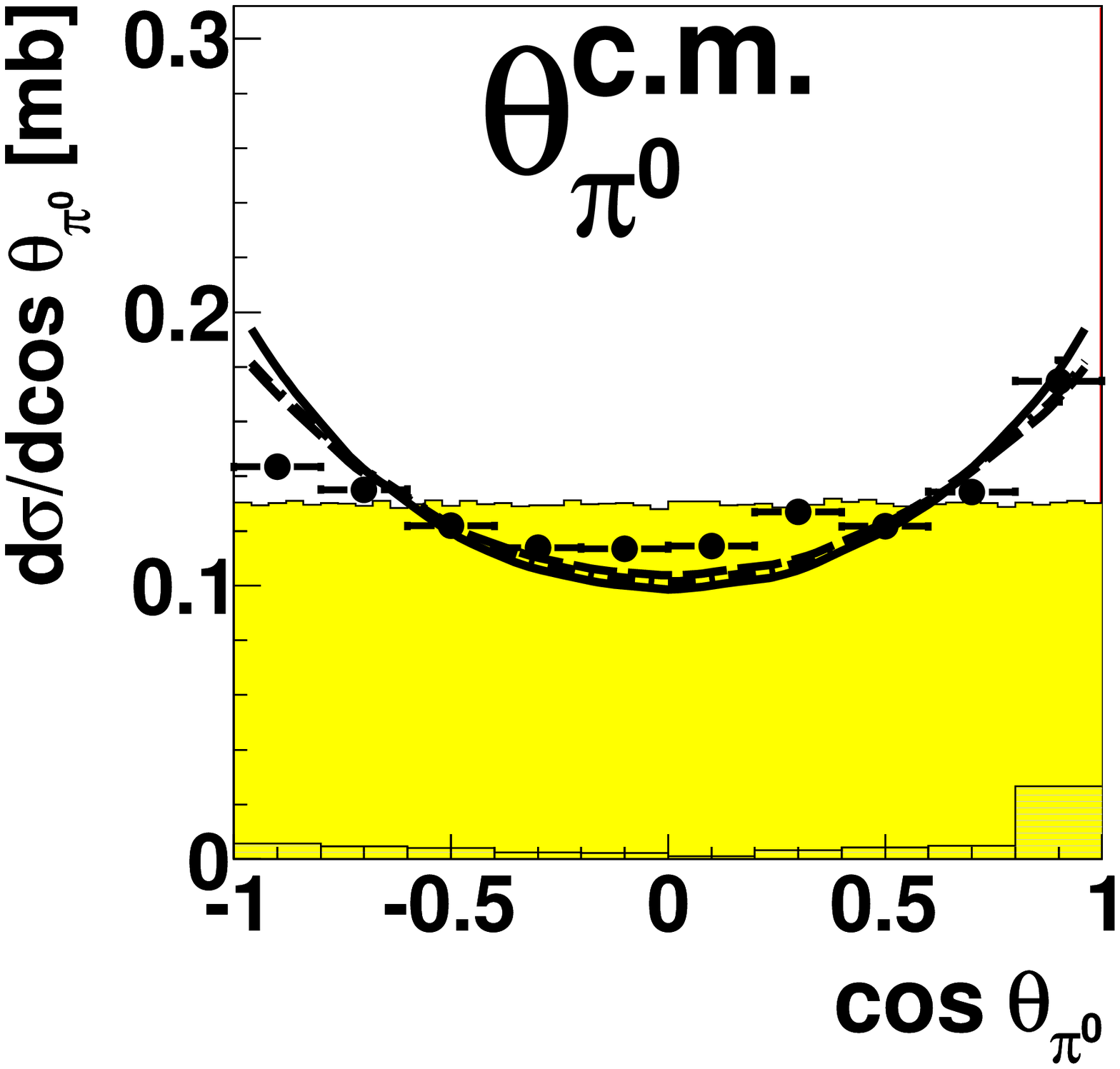}
\caption{(Color online) 
  Same as Fig. 3 but for the 
   distributions of the cm angles $\Theta_p^{c.m.}$ (top) and
   $\Theta_{\pi^0}^{c.m.}$ (bottom). 
}
\label{fig6}
\end{center}
\end{figure}
%%%%%%%%%%%%%%%%%%%%%%%%%%%%%%%%%%%%%%%%%%%%%%%%%
%\begin{figure}
%\begin{center}
%\includegraphics[width=0.4\textwidth]{invp2pi0.eps}
%\includegraphics[width=0.4\textwidth]{nppi0pi0_cos2pi0.eps}
%\includegraphics[width=0.4\textwidth]{nppi0pi0_cosn.eps}
%\caption{(Color online) 
%  Not to be included in the paper. 
%}
%\label{figx}
%\end{center}
%\end{figure}
%%%%%%%%%%%%%%%%%%%%%%%%%%%%%%%%%

All measured differential distributions are markedly different in shape from
pure phase space distributions (shaded areas in Figs. 3 - 6), but 
% with the exception of the $M_{\pi^0\pi^0}$ spectrum
close to the predictions %of the modified Valencia calculations 
both with (dashed and dash-dotted lines) and without (solid lines) inclusion
of the $d^*$ resonance. 
 
The invariant mass spectra for $M_{p\pi^0}$, $M_{n\pi^0}$, $M_{n\pi^0\pi^0}$
and $M_{pn\pi^0}$ (Figs. 4 - 5) are characterized by $\Delta$ and $N\Delta$
dynamics as they naturally appear in the deexcitation process of an
intermediate $\Delta\Delta$ system created either by $d^*$ decay or via
$t$-channel meson exchange.  

The pion angular distribution (Fig. 6) behaves as expected from the $p$-wave
decay of the $\Delta$ resonance. And also the proton angular distribution is
similarly curved. Both t-channel meson exchange and the $J^P = 3^+$ requirement
for $d^*$ formation predict comparable shapes in agreement with the data. 

The $M_{pn}$ and $M_{\pi^0\pi^0}$ spectra (Fig. 3) need a more thorough 
discussion. The data of the $M_{\pi^0\pi^0}$ spectrum appear to be quite well
described by the calculations, which hardly deviate from each other. At small
invariant masses though, in the range 0.3 - 0.4 GeV/c$^2$, there is an
indication of a small surplus of strength. Taken the uncertainties inherent 
in the data and in the theoretical description, these deviations appear not to
be particularly significant. Therefore, if this
constitutes a sign of the ABC effect, then it is obviously very small in
this reaction. Note that contrary to the situation in the $pn \to pp\pi^0\pi^-$
reaction, where the pion pair has to be in relative $p$-wave and hence the
ABC-effect is absent, the pion pair here is preferentially in relative
$s$-wave allowing thus, in principle, the occurrence of the ABC effect. 
Hence, the finding that there is no or nearly no ABC effect comes as a
surprise at least for some of its interpretations. This finding is of no
surprise, if the ABC effect is described by a formfactor at the $\Delta\Delta$
vertex of the $d^*$ decay \cite{prl2011}. However, then a problem arises with
the description of the $M_{pn}$ spectrum, as we discuss in the following.

The $M_{pn}$ spectrum peaks sharply at its low-mass threshold,
which is characteristic for a strong $np$ FSI as discussed above. This
low-mass peaking is well accounted for by the modified Valencia
calculations (solid lines in Figs. 3 - 6) . Inclusion of the $d^*$ resonance
as outlined in Ref. \cite{prl2011}
(dashed lines) exaggerates the low-mass peaking deteriorating thus the
agreement with the data. The reason 
for this behavior is the formfactor at the $\Delta\Delta$ decay vertex of $d^*$
introduced in Ref. \cite{prl2011} for the description of the ABC effect, {\it
  i.e.} the low-mass enhancement in the $M_{(\pi\pi)^0}$ spectra observed in
double-pionic fusion reactions. However, as already pointed out in 
Ref.~\cite{pp0-}, this formfactor acts only on the $M_{\pi^0\pi^0}$ and
$M_{\pi^+\pi^-}$ spectra, 
if the nucleon pair is bound in a final nuclear system. If this is not the
case, then the formfactor acts predominantly on the invariant-mass spectrum of
the nucleon pair. This is illustrated by comparison of the calculations
including $d^*$ with (dashed) and without (dash-dotted) this formfactor. As we
see, the formfactor hardly changes the $M_{\pi^0\pi^0}$ distribution, but
shuffles substantial strength in the $M_{pn}$ spectrum to low masses -- thus
overshooting the observed low-mass enhancement.

This finding indicates that the formfactor introduced in Ref.~\cite{prl2011} on
purely phenomenological grounds for the description of the ABC effect is
possibly at
variance with the data for isoscalar two-pion production in non-fusion
channels. Hence alternative solutions for this phenomenon may have to be looked
for, such as $d$-wave contributions in the intermediate
$\Delta\Delta$ system and/or final nucleon-pair \cite{Zhang,Huang}.

Another alternative involving $d$-waves has been proposed recently by
Platinova and Kukulin \cite{kuk}. In their ansatz they  
assume the $d^*$ resonance not only to decay into the $d\pi^0\pi^0$ channel
via the route $d^* \to \Delta^+\Delta^0 \to d\pi^0\pi^0$ \footnote{actually
  they consider the decay $d^* \to D_{12}^{++}\pi^0 \to d\pi^0\pi^0$ with
  $D_{12}^{++}$ being a $I(J^P) = 1(2^+)$ state near the $N\Delta$ threshold,
  but since the pion emitted in the $d^*$ decay is in relative $p$-wave to
  $D_{12}$, this route is practically indistinguishable from a $d^* \to
  \Delta^+\Delta^0$ decay at the given kinematic conditions}, but also via the
route $d^*
\to d\sigma \to d\pi^0\pi^0$. Since $\sigma$ is a spin zero object, it has to
be in relative $d$-wave to the deuteron in this decay process, in order to
satisfy the resonance condition of $J^P~=~3^+$. In consequence the available
momentum in this decay process is concentrated in the relative motion between
$d$ and $\sigma$ leaving thus only small relative momenta between the two
emerging
pions. Therefore the $M_{\pi^0\pi^0}$ distribution is expected to be peaked at
low masses -- {\it i.e.}, the low-mass enhancement (ABC effect) in this model is
made by the $d\sigma$ decay branch (in the amount of about 5$\%$) and not by a
formfactor as introduced in 
Ref. \cite{prl2011}.  The enhancement in this model is further increased by
interference of the $d\sigma$ decay amplitude with the decay amplitude via the
$\Delta^+\Delta^0$ system. It appears straightforward to extend this ansatz also
to reaction channels, where the $np$ system is unbound. 
%The dash-dotted curves
%in Figs. 4 - 6 show the results with this ansatz, which gives a practically
%perfect description of all differential distributions. 
However, since we hardly observe a low-mass enhancement (ABC effect) in the
$M_{\pi^0\pi^0}$ spectrum, much less $d^* \to d\sigma$ contribution is needed
here than in the $pn \to d\pi^0\pi^0$ reaction -- which possibly poses a
consistency problem for this ansatz \cite{kuk}.

Another point of concern with this ansatz is that mass and width of the sigma
meson have been fitted to the $pn \to d\pi^0\pi^0$ data in Ref. \cite{kuk} with the
result that $m_{\sigma} \approx$ 300 MeV and $\Gamma_{\sigma} \approx$ 100
MeV.  Both values are much smaller than the generally accepted values for the
sigma meson \cite{PDG}, which are $m_{\sigma}$ = (400 - 550) MeV and
$\Gamma_{\sigma}$ = (400 - 700) MeV. In Ref.~\cite{kuk} it has been argued
that these deviations could be a sign of chiral restoration in the
hadronic/nuclear environment - in particular within the six-quark
bag. However, any evidence for this hypothesis from other experiments is lacking
so far.  
%If true, then the ABC enhancement should be even more
%concentrated towards the $M_{\pi\pi}$ low-mass threshold in case of the
%double-pionic fusion to $^4$He, since the mass density in the interior of
%$^4$He is very high. In fact, that is what is observed experimentally
%\cite{AP}.  
Whether the enhanced ABC effect observed in the double-pionic fusion to $^4$He
\cite{AP} is in support of such an argumentation is an open question.

\section{Conclusions}

The $np \to np\pi^0\pi^0$ reaction, for which no dedicated previous data exist, has been investigated by exclusive and kinematically complete
measurements. They have been  
carried out in quasifree kinematics with a deuteron beam impinging on a hydrogen
pellet target. Utilizing the nucleons' Fermi motion in the deuteron projectile
an energy region of 2.35 GeV $< \sqrt s <$ 2.41 GeV could be covered
corresponding to an incident lab energy range of 1.07 - 1.23 GeV. This
energy region covers the region of the $d^*$ resonance. The data are in
agreement with a resonance contribution of about 200 $\mu$b,
as predicted by F\"aldt and Wilkin \cite{col} as well as by Albaladejo and
Oset \cite{oset}. 

In general, the differential data are reasonably well described by
calculations, which include both the $d^*$ resonance and the conventional
$t$-channel processes. 

The data indicate only a very small low-mass enhancement (ABC effect) in the
$\pi^0\pi^0$-invariant mass distribution. Though this not in disagreement with
the phenomenological ansatz of a formfactor at the $d^* \to \Delta\Delta$ decay
vertex introduced in Ref.~\cite{prl2011}, the worsening of the description of
the $M_{pn}$ spectrum by use of this formfactor calls possibly for an improved
explanation of the ABC effect in connection with the $d^*$ resonance.

After having found evidences for the $d^*$ resonance in the
$d\pi^0\pi^0$, $d\pi^+\pi^-$ and $ pp\pi^0\pi^-$ channels, the channel
investigated here has been one of the two remaining two-pion production
channels, where the predicted contributions of the $d^*$ resonance had not yet
been checked experimentally. As we have shown now, the data for the
$np\pi^0\pi^0$ channel are consistent with the $d^*$ hypothesis and provide an
experimentally determined branching for the $d^*$ decay into this
channel. 

Since $d^*$ has been observed meanwhile also in the elastic channel by
polarized $\vec{n}p$ scattering, the only remaining unexplored decay channel is
$np\pi^+\pi^-$. This channel has been measured recently at HADES and
preliminary results have been presented already at conferences
\cite{kuril,agaki,prag}. It will be highly interesting, not only to obtain
total cross sections for this channel, but also differential distributions. Of
particular interest will be  the $M_{pn}$ and $M_{\pi^+\pi^-}$ distributions
as discussed in this work.

\section{Acknowledgments}

We acknowledge valuable discussions with  
V. Kukulin, E. Oset and C. Wilkin on this issue. We are particularly indebted to
L. Alvarez-Ruso for using his code.  
This work has been supported by Forschungszentrum J\"ulich (COSY-FFE), DFG, 
the Foundation for Polish Science through the MPD programme 
and by the Polish National Science Centre through the Grants No.
2011/01/B/ST2/00431 and 2013/11/N/ST2/04152.

\end{document}